# Statistical Evaluation of the Azimuth and Elevation Angles Seen at the Output of the Receiving Antenna

Cezary Ziółkowski and Jan M. Kelner

*Abstract*—**A method to evaluate the statistical properties of the reception angle seen at the input receiver that considers the receiving antenna pattern is presented. In particular, the impact of the direction and beamwidth of the antenna pattern on distribution of the reception angle is shown on the basis of 3D simulation studies. The obtained results show significant differences between distributions of angle of arrival and angle of reception. This means that the presented new method allows assessing the impact of the receiving antenna pattern on the correlation and spectral characteristics at the receiver input in simulation studies of wireless channel. The use of this method also provides an opportunity for analysis of a co-existence between small cells and wireless backhaul, what is currently a significant problem in designing 5G networks.**

*Index Terms*—**Angle of arrival, angle of reception, angle spread, antenna radiation pattern, azimuthal and elevational planes, channel models, channel modeling, directional receiving antenna, geometric channel models, half power beamwidth.**

## I. Introduction

The direction and spatial shape of the pattern of the receiving antenna significantly affect the statistical properties of the signal reception angle. These properties that are described by probability density function (PDF) deform the correlation and spectral characteristics of the signals transmitted in wireless channels [1-3]. Therefore, mapping the spatial distribution of the reception angle "seen at the output of the receiving antenna", is required to obtain the convergence of the simulation studies and actual measurements. The use of this method also provides an opportunity for analysis of a co-existence between small cells and wireless backhaul (WB). Especially, it will allow to evaluate the interference caused by 5G access point towards WB receiver needed to determine a minimum distance between 5G and WB deployments. This current and significant problem in designing 5G networks stems from narrowbeam antenna patterns, which ensure minimizing power consumption and increase the range of radio links.

The directivity of the receiving antenna results in a spatial selection of propagation paths. Therefore, the signal arriving at the input of the receiver is a superposition of signals from all propagation paths and their levels are formed by the receiving antenna pattern. This is the cause of differentiation of the statistical properties of the angle of arrival (AOA) in the surroundings of the receiving antenna and the angle of reception (AOR) that is "seen at the output of the receiving antenna". In the case of an omnidirectional antenna in the azimuth plane and its large beamwidth in the elevation plane, PDF of AOA and PDF of AOR are convergent. In literature, we can find many models and mapping methods of the statistical properties of angle that consider the scattering phenomenon both on the azimuth and elevation planes [4-8]. However, these models and methods focus only on mapping AOA and are mainly based on omnidirectional antennas. Only a few of them consider the sector antennas but merely on the transmission side and in a simplified manner [9,10]. Therefore, these models and methods can be used in simulation only in scenarios where PDFs of AOA and AOR are convergent. For the sectoral and narrowbeam antennas, the effect consideration of spatial filtering antenna is required to evaluate the correlation and spectral signal properties at the input of the receiver. 3GPP channel model gives such possibilities but only to a limited scope [11]. These limitations are the result of using only a few strictly defined scenarios that specify the parameters of propagation phenomena. The solution that is presented in this communication, addresses this problem and is applicable in channel simulation studies and in empirical data analysis. In this case, the developed method enables to use any propagation scenario that is defined by the power delay profile (PDP) or power delay spectrum (PDS). In addition, the consideration of the transmitter and receiver antenna patterns in the angular power distribution is an innovative contribution of this paper.

Here, we present a method to assess the statistical properties of AOR that consider the impact of the receiving antenna pattern on the direction of signal reception in 3D, and the beamwidth of the antenna pattern on PDF of AOR is shown for the simulation scenarios, whose parameters are defined on the basis of the 3GPP channel model [11]. The remainder of this communication is organized as follows. The system geometry and 3D channel model for AOA generation is presented in Section II. Determination of AOR and estimation of its PDF is described in Section III. The next section includes the results of the simulation studies that show the effects of the direction and beamwidth of the antenna pattern on spread of AOR. Section IV provides some concluding remarks.

## II. 3D Channel Model

The geometrical model of the channel is used to generate sets of AOAs that describe the space scattering of signals as a set of half-ellipsoids. The number and spatial parameters of these half-ellipsoids are defined on the basis of the power delay profile (PDP) or power delay spectrum (PDS). The local extremes of these characteristics represent signal components that arrive at the receiver with the same delay and form so-called time clusters. The amount and the position in time domain of these extremes define the number and size of the individual half-ellipsoids. The geometry of the channel model is shown in Fig. 1.

Adopted geometry is a basis to generate the angular parameters of the propagation paths arriving at the receiver with a delay relative to the direct path. These parameters are sets of angles in the azimuth and elevation planes, and the levels of power that characterize the intensity of each path. In addition to the parameters of the paths that represent delayed components of the signal, the parameters of the local scattering paths are generated. In this case, the von Mises PDF is used. As a result, we obtain the sets of the angles in the elevation $\Theta$ and azimuth $\Phi$ planes, and power levels $\mathbf{P}$ that describe each propagation path arriving at the receiver.

$$\mathbf{\Theta} = \left\{ \theta_{ij} \right\}_{i=0,\, j=1}^{N,M_i}, \quad \mathbf{\Phi} = \left\{ \varphi_{ij} \right\}_{i=0,\, j=1}^{N,M_i}, \quad \mathbf{P} = \left\{ p_{ij} \right\}_{i=0,\, j=1}^{N,M_i} \quad (1)$$

where: $i$ is the number of the time cluster (half-ellipsoid), $j$ is the number of component in the $i$th time cluster, $N$ represents the number all time clusters (half-ellipsoids), and $M_i$ means the number of the components (propagation paths) in the $i$th time cluster. The elements with $i = 0$ represent the local scattering components.

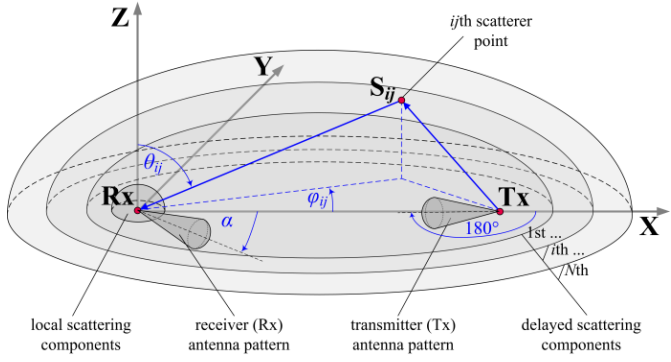

Fig. 1.   Geometry of 3D channel model.

This channel model is one of a few that considers the impact of radiation transmitter antenna on the statistical properties of AOA. A detailed description of the generation procedure of $\mathbf{\Theta}$, $\mathbf{\Phi}$, and $\mathbf{P}$ is presented in [12]. In this publication, the analysis of the statistical properties of the generated sets of the angles shows the compliance with the measurement data. This justifies the use of this model to determine the angle and power parameters of the received signal components in simulation studies of channel. The used geometric channel model is an extension of our previous works, i.a., [13,14].

## III. ESTIMATION OF AOR DISTRIBUTION

Evaluation of the influence of the receiving antenna at AOR is based on the input data that are $\mathbf{\Theta}$, $\mathbf{\Phi}$, $\mathbf{P}$, and the power pattern of this antenna, $g_R^2(\theta_{ij}, \varphi_{ij})$. The sets of the angles and powers that represent the parameters of the propagation paths, can be written in integrated form as $\left\{ p_{ij}(\theta_{ij}, \varphi_{ij}) \right\}_{i=0,\, j=1}^{N,M_i}$. This form shows that each element of this set represents the power that arrives at the receiver from $(\theta_{ij}, \varphi_{ij})$ direction. Thus, for the $j$th path of the $i$th half-ellipsoid, the signal power, $p_{Rij}(\theta_{ij}, \varphi_{ij})$, at the output of the receiving antenna expresses the following relationship:

$$p_{Rij}(\theta_{ij}, \varphi_{ij}) = p_{ij}(\theta_{ij}, \varphi_{ij}) g_R^2(\theta_{ij}, \varphi_{ij}) \quad (2)$$

In simulation studies, the Gaussian beam is commonly parametrized power pattern of the antennas [15,16]

$$g_R^2(\theta, \varphi) = G_R \exp\left( -\frac{(\pi/2 - \theta)^2}{\sigma_{g\theta}^2} \right) \exp\left( -\frac{\varphi^2}{\sigma_{g\varphi}^2} \right) \quad (3)$$

where: $G_R$ means the boresight gain of the antenna, $\sigma_{g\theta}$ and $\sigma_{g\varphi}$ represent beamwidths of the antenna pattern in the elevation and azimuth planes, respectively. These parameters are closely related to the half power beamwidths (HPBWs) in the respective planes (see [15]).

These relationships are the basis for the transformation of the signal power from the surroundings of the antenna to the input of the receiver.

Let $\mathbf{O}(\theta, \varphi) = \left\{ (i, j) : \theta_{ij} \in (\theta \pm \varepsilon_\theta) \wedge \varphi_{ij} \in (\varphi \pm \varepsilon_\varphi) \right\}$, where $\varepsilon_\theta$ and $\varepsilon_\varphi$ are the neighborhoods of $\theta$ and $\varphi$, respectively. Thus, $\sum_{\mathbf{O}(\theta,\varphi)} p_{Rij}(\theta_{ij}, \varphi_{ij})$ represents the total power of the signal that arrives at the input of the receiver from $(\theta \pm \varepsilon_\theta, \varphi \pm \varepsilon_\varphi)$ sector. Let $Q$ means the number of all propagation paths reaching the receiver and let in the neighborhood of ever $\theta$ and $\varphi$ is at least one angle that describes the path. Then, for $Q \to \infty$, the size of the neighborhood approaches zero for each angle ($\varepsilon_\theta \to 0$ and $\varepsilon_\varphi \to 0$). Consequently, each sum related to $\varepsilon_\theta$ and $\varepsilon_\varphi$ represents the power that falls on the elementary interval of the angle. Therefore, in the limit, we obtain the power angle spectrum, $P_R(\theta, \varphi)$

$$\frac{\sum_{\mathbf{O}(\theta,\varphi)} p_{Rij}(\theta_{ij}, \varphi_{ij})}{4\varepsilon_\theta \varepsilon_\varphi} \xrightarrow[\substack{Q \to \infty \\ (\varepsilon_\theta \to 0 \,\wedge\, \varepsilon_\varphi \to 0)}]{} P_R(\theta, \varphi) \quad (4)$$

It means that these finite sums can be treated as an estimator of $P_R(\theta, \varphi)$. Note also that the signal average power, $P_0$, at the input of the receiver is estimated by

$$\frac{\sum_{i=0}^{N} \sum_{j=1}^{M_i} p_{Rij}(\theta_{ij}, \varphi_{ij})}{4\varepsilon_\theta \varepsilon_\varphi} \xrightarrow[\substack{Q \to \infty \\ (\varepsilon_\theta \to 0 \,\wedge\, \varepsilon_\varphi \to 0)}]{} P_0 \quad (5)$$

In practice, we can present $P_R(\theta, \varphi)$ in the form

$$P_R(\theta, \varphi) = P_0 f_R(\theta, \varphi) \quad (6)$$

where $f_R(\theta, \varphi)$ means PDF of AOR. On the basis of (6), the estimator of PDF of AOR, $\tilde{f}_R(\theta, \varphi)$, is

$$\tilde{f}_R(\theta, \varphi) = C_0 \frac{\sum_{\mathbf{O}(\theta,\varphi)} p_{Rij}(\theta_{ij}, \varphi_{ij})}{\sum_{i=0}^{N} \sum_{j=1}^{M_i} p_{Rij}(\theta_{ij}, \varphi_{ij})} \quad (7)$$

where $C_0$ is a normalizing constant that is associated with $\varepsilon_\theta$, $\varepsilon_\varphi$, and provides a condition $\lim_{\substack{\varepsilon_\theta \to 0 \\ \varepsilon_\varphi \to 0}} \int_0^{90°} \int_{-180°}^{180°} \tilde{f}_R(\theta, \varphi) \mathrm{d}\theta \, \mathrm{d}\varphi = 1$.

In the real environment, scattering phenomena that occur in the elevation and azimuth planes are independent. Hence, marginal PDFs of AOR in the elevation and azimuth planes can be represented in the following forms:

$$\tilde{f}_R(\theta) = C_\theta \frac{\sum_{\mathbf{K}(\theta)} p_{Rij}(\theta_{ij})}{\sum_{i=0}^{N} \sum_{j=1}^{M_i} p_{Rij}(\theta_{ij})}, \quad \tilde{f}_R(\phi) = C_\phi \frac{\sum_{\mathbf{L}(\phi)} p_{Rij}(\phi_{ij})}{\sum_{i=0}^{N} \sum_{j=1}^{M_i} p_{Rij}(\phi_{ij})} \quad (8)$$

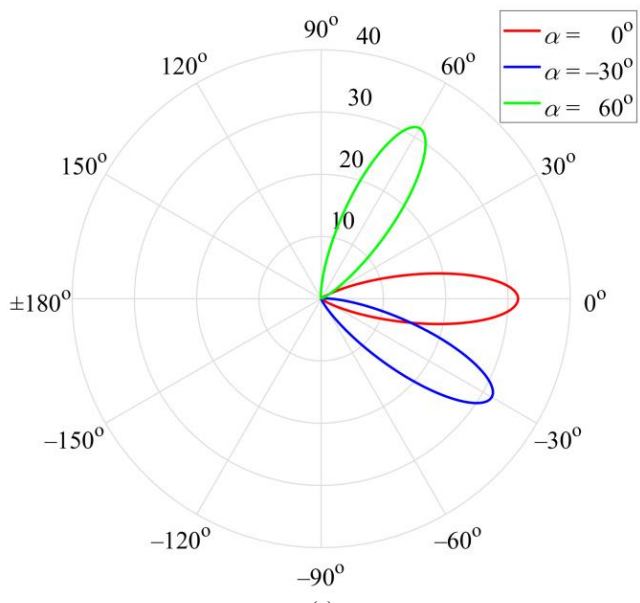

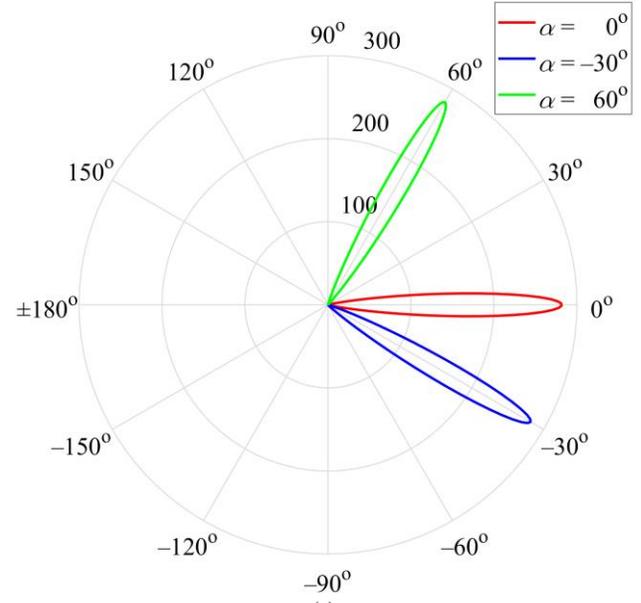

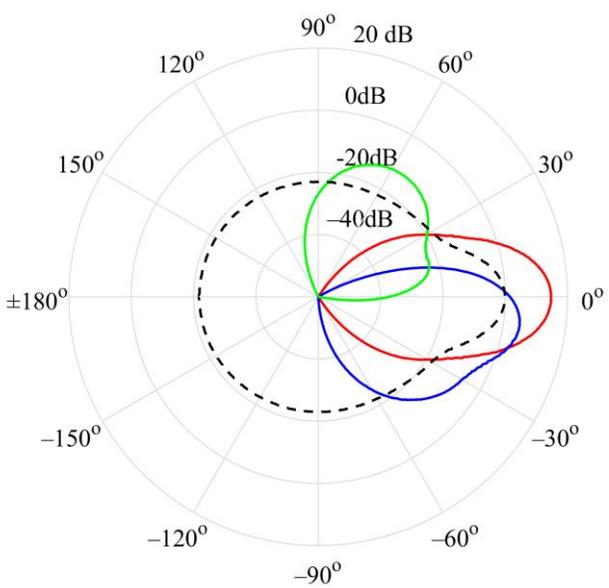

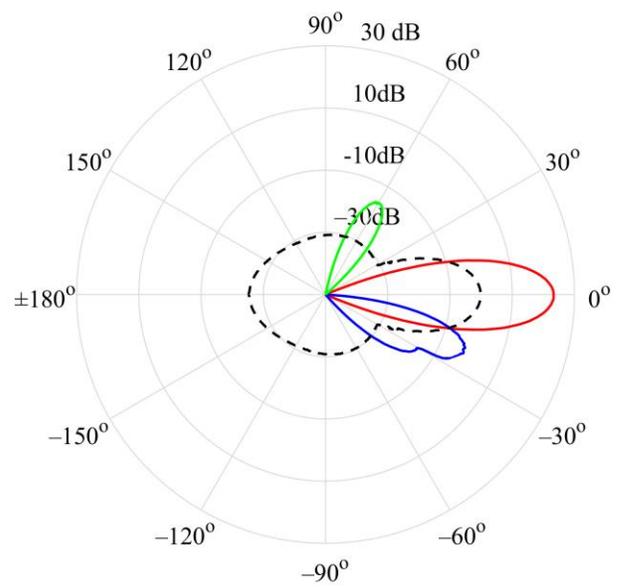

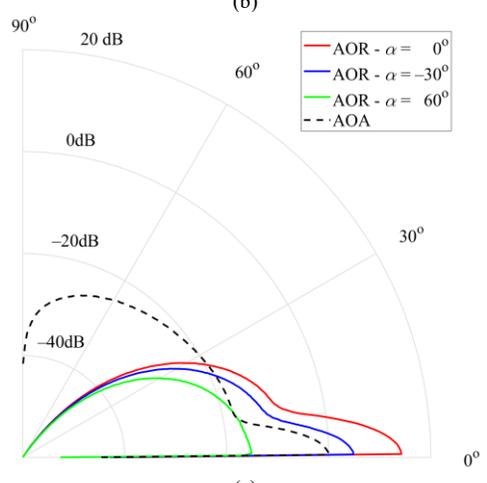

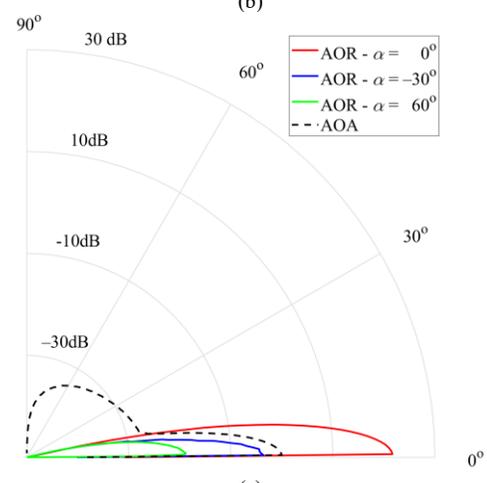

Fig. 2. (a) Power pattern of widebeam antenna for $\alpha = -30°$, $0°$, $60°$ in azimuth plane (linear scale), (b) $P_R(\varphi)$, and (c) $P_R(\theta)$ for scenario 2.

Fig. 3. (a) Power pattern of narrowbeam antenna for $\alpha = -30°$, $0°$, $60°$ in azimuth plane (linear scale), (b) $P_R(\varphi)$, and (c) $P_R(\theta)$ for scenario 2.

where $\widetilde{f}_R(\theta)$ and $\widetilde{f}_R(\varphi)$ are the estimators of PDFs of AOR in the elevation and azimuth planes, respectively, $\mathbf{K}(\theta) = \{(i,j) : \theta_{ij} \in (\theta \pm \varepsilon_\theta)\}$ , $\mathbf{L}(\varphi) = \{(i,j) : \varphi_{ij} \in (\varphi \pm \varepsilon_\varphi)\}$ , whereas $C_\theta$ and $C_\varphi$ fulfill conditions $\lim\limits_{\varepsilon_\theta \to 0} \int\limits_0^{90^\circ} \widetilde{f}_R(\theta) \mathrm{d}\theta = 1$ and $\lim\limits_{\varepsilon_\varphi \to 0} \int\limits_{-180^\circ}^{180^\circ} \widetilde{f}_R(\varphi) \mathrm{d}\varphi = 1$ .

In simulation and empirical studies of channel, (2) and (8) are the basis for the practical assessment of the impact of the receiving antenna pattern on the statistical properties of AOR.

## IV. ANTENNA PATTERN PARAMETERS AND PDF OF AOR

In this section, we show the impact of the receiving antenna on the statistical properties of AOR that is viewed from the receiver input. Here, we focus on the changes of PDF of AOR that result from changes in both the direction, $\alpha$ , and beamwidths, $HPBW_{\theta,\varphi}$, of the transmitting and receiving antenna patterns. To include these relations we introduce following designations in the analytical description of PDFs: $\widetilde{f}_R(\theta) \equiv \widetilde{f}_{R\Omega}(\theta)$ and $\widetilde{f}_R(\varphi) \equiv \widetilde{f}_{R\Omega}(\varphi)$ , where $\Omega = (\alpha, HPBW_{\theta,\varphi})$. In the simulation studies, PDPs from 3GPP non-line-of-sight propagation scenarios [11, Table 7.7.2-2] are assumed to design the geometric channel model. As an example, we used three urban macro (UMa) scenarios for 28 GHz. The scenarios 1, 2, and 3 are defined as short-, normal-, and long-delay profiles [11, Table 7.7.3-2] that represent the average results of measurement campaigns in diverse urban environments. All tests are carried out at narrowbeam and widebeam antennas for 28 GHz whose parameters are as follows: the widebeam antenna − $G_R = 15.0 \, \mathrm{dBi}$ , $HPBW_\theta = 30.0^\circ$ , and $HPBW_\varphi = 28.8^\circ$ , the narrowbeam antenna − $G_R = 24.5 \, \mathrm{dBi}$ , $HPBW_\theta = 8.6^\circ$ , and $HPBW_\varphi = 10.9^\circ$ , respectively. These parameters are adopted on the basis of measurement campaigns [16-18], which aim was to assess the propagation phenomena in wireless links of 5G networks. To evaluate 3D modeling procedure, 200 Monte Carlo runs were carried out in the Matlab.

In the first step, we consider the impact of $\alpha$ that is the angle between the signal source and receiving antenna pattern direction in the azimuth plane. Based on (6) and (8), the graphs of $P_R(\theta,\varphi)$ in the azimuth, $P_R(\varphi)$ , and elevation, $P_R(\theta)$ , planes for the widebeam and narrowbeam antennas, and scenario 2 are presented in Figs. 2 and 3, respectively. In each of the figures, a case of AOA is considered to assess the impact of $g_R^2(\theta_{ij}, \varphi_{ij})$ on $P_R(\theta)$ and $P_R(\varphi)$ . In addition, the angular positions of the receiving antenna patterns are shown in Figs. 2 a) and 3 a), respectively. For the widebeam antenna, the increase of $|\alpha|$ to 120° causes a decrease of 30 dB and 27 dB in the maximum of $P_R(\theta)$ and $P_R(\varphi)$ . Whereas for the narrowbeam antenna, this reduction is 46 dB and 40 dB, respectively.

As we can see, the change in $\alpha$ is reflected not only in level changes, but also in the changes of the angular dispersion of the receiving signal power. For scenario 2, the influence of $\alpha$ on the statistical properties of the power dispersion that represent PDFs, are illustrated in Fig. 4.

As can be seen in Fig. 4 b), PDFs are asymmetrical for $\alpha \neq 0$ , which is clearly visible with increasing $|\alpha|$. To quantify the influence of $\alpha$ on the intensity of the AOR scattering phenomenon, the standard deviations are used. For the elevation and azimuth planes, these parameters are determined, respectively, as

$$\sigma_\theta(\Omega) = \sqrt{\frac{1}{J}\sum_{n=1}^{J}\theta_n^2 \widetilde{f}_{R\Omega}(\theta_n) - \left(\frac{1}{J}\sum_{n=1}^{J}\theta_n \widetilde{f}_{R\Omega}(\theta_n)\right)^2} \quad (9)$$

$$\sigma_\varphi(\Omega) = \sqrt{\frac{1}{J}\sum_{n=1}^{J}\varphi_n^2 \widetilde{f}_{R\Omega}(\varphi_n) - \left(\frac{1}{J}\sum_{n=1}^{J}\varphi_n \widetilde{f}_{R\Omega}(\varphi_n)\right)^2} \quad (10)$$

where $J$ is the number of data from the simulation results.

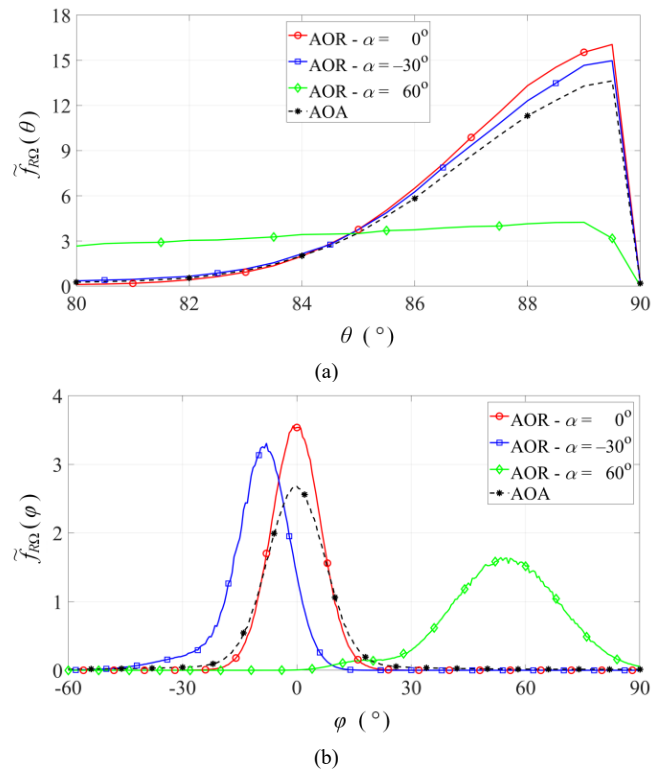

(a)

(b)

Fig. 4. PDFs of AOR for $\alpha = -30^\circ, 0^\circ, 60^\circ$ in (a) elevation and (b) azimuth planes.

For the widebeam and narrowbeam antennas, the graphs of $\sigma_\theta(\Omega)$ and $\sigma_\varphi(\Omega)$ versus $|\alpha|$ are shown in Fig. 5. For the widebeam and narrowbeam transmitting antennas, the standard deviations of AOA are ( $\sigma_\theta = 14^\circ$ , $\sigma_\varphi = 34^\circ$ ), and ( $\sigma_\theta = 6^\circ$ , $\sigma_\varphi = 19^\circ$ ), respectively. Fig. 5 shows that $\sigma_\theta(\Omega)$ and $\sigma_\varphi(\Omega)$ reach minimum for $\alpha = 0$ . It means that the beamwidth of the receiving antenna limits the intensity of the power dispersion by approximately 11°, 27° and 4°, 15° compared to AOA for the widebeam and narrowbeam antenna, respectively. For $|\alpha| \geq 30^\circ$ , the scattering intensity of AOR significantly increases particularly in the azimuth plane.

The relationships (2) and (8) also provide an opportunity to analyze the changes in PDF of AOR as a function of the antenna pattern beamwidth. In this case, HPBW is analyzed as a parameter of PDFs for angles in the elevation and azimuth planes. The example of mapping the impact of the antenna pattern beamwidth on PDF of AOR is considered for $\alpha = 0$. For the widebeam and narrowbeam antennas, the graphs of PDFs of AOR in the elevation and azimuth planes are shown in Figs. 6 a) and 6 b), respectively. Additionally, PDFs of AOA at the surrounding the receiving antenna with regarding the transmitting antenna pattern are presented to compare the statistical properties of the reception angle.

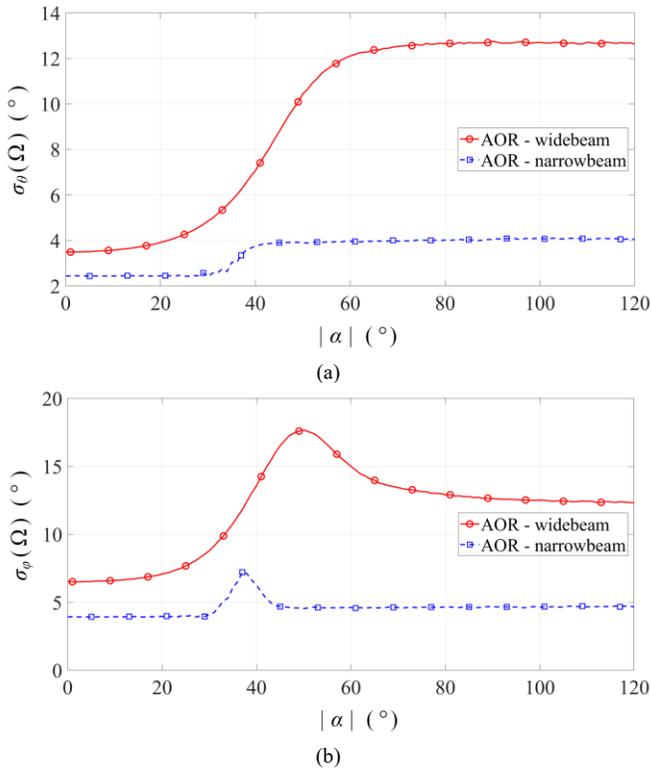

Fig. 5. Graphs of $\sigma_\theta(\Omega)$ and $\sigma_\phi(\Omega)$ versus $|\alpha|$ for widebeam and narrowbeam antennas in (a) elevation and (b) azimuth planes.

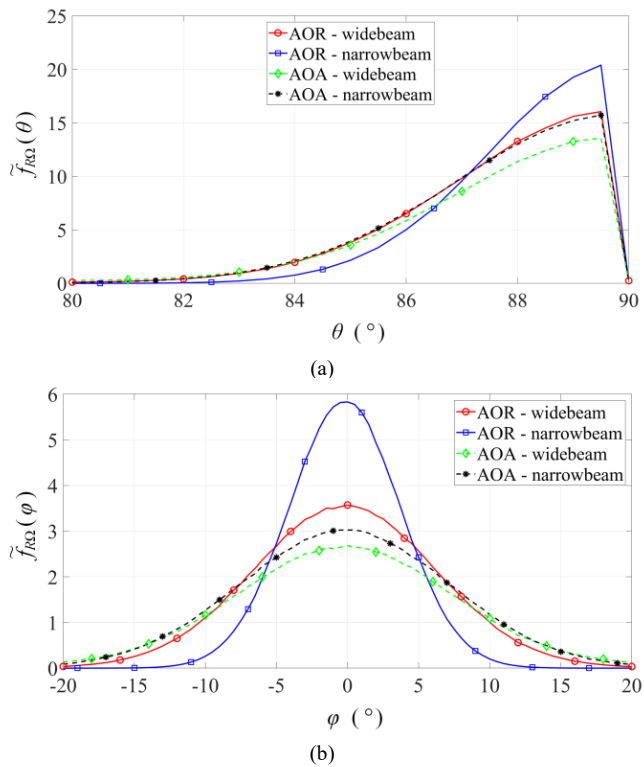

Fig. 6. PDFs of AOR and AOA for widebeam and narrowbeam antennas in (a) elevation and (b) azimuth planes.

It is obvious, that the AOR dispersion is reduced along with a decrease in the pattern beamwidth of the receiving antenna. However, quantitative assessment of the effect is possible because of the analysis of the simulation data that is presented in this communication. Here, we limit this analysis only to assess the impact of changes in the pattern beamwidth in the azimuth plane. In Fig. 7, $\sigma_\phi(\Omega)$ is shown as a function of $HPBW_\phi$ for three UMa scenarios. Simulation test are performed on the assumptions $\alpha = 0$ and $HPBW_\theta = 30°$.

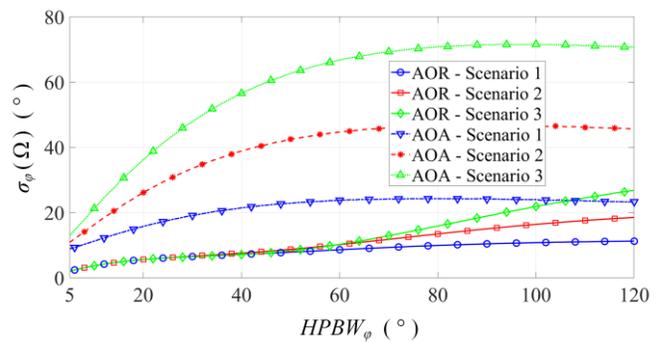

Fig. 7. Graphs of $\sigma_\phi(\Omega)$ versus $HPBW_\phi$ for UMa scenarios ($\alpha = 0$, $HPBW_\theta = 30°$).

Based on the graphs of $\sigma_\phi(\Omega)$ for AOA, we can see that the type of environment significantly differentiated the intensity of the scattering phenomenon. This environmental impact is already present when the $HPBW_\phi$ of the transmitting antenna exceeds 5°. Applying the narrowbeam pattern of the receiving antenna substantially reduces the impact of the environment on the AOR distribution. In this case, the environmental impact appears only for $HPBW_\phi > 40°$ and it is insignificant compared to AOA.

## V. CONCLUSION

This communication presents methods to map and assess the impact of the receiving antenna on the statistical properties of AOR. In the case of using the receiving sector antennas, the obtained results show significant differences between PDFs of AOA and AOR. This means that in both simulation and empirical studies of channel, the assessment of the correlation and spectral characteristics of the signal at the receiver input requires consideration of the receiving antenna pattern. The modeling procedure presented in this communication, is a new method of mapping the statistical characteristics of AOR that considers the receiving antenna parameters and type of propagation environment. The use of this processing data in simulation studies of wireless system channels significantly reduce the approximation error of the modeling results with respect to actual measurements of signal at the output of the receiving antenna. The developed method gives us the opportunity to analyze the co-existence between small cells and WB, what is currently a significant problem in designing 5G networks. The significance of this problem is presented in [19]. In this case, the 3GPP [11] model was used, which defines determined spatial scenarios and considers antenna patterns in a simple filtering procedure. The use of our method gives the opportunity to consider the actual environmental conditions that are defined by measurement PDPs.